\documentclass[
reprint,           
superscriptaddress,
amsmath,           
amssymb,           
aps,               
prl,               
notitlepage,       
floatfix,          
nofootinbib,
]{revtex4-1}

\usepackage{tensor}     
\usepackage{graphicx}   
\usepackage[
colorlinks=true,        
citecolor=blue,         
linkcolor=blue,         
urlcolor=blue           
]{hyperref}             
\usepackage{bm}         
\usepackage{xcolor}     
\usepackage{tabulary}
\usepackage{color}      
\usepackage[utf8]{inputenc} 
\usepackage[section]{placeins} 

\newcommand{\nc}{\newcommand*} 

\nc{\al}{\alpha}
\nc{\s}{\sigma}
\nc{\dt}{\delta}
\nc{\Dt}{\Delta}
\nc{\Ld}{\Lambda}
\nc{\p}{\partial}
\nc{\om}{\omega}
\nc{\Om}{\Omega}
\nc{\rd}{\mathrm{d}}
\nc{\Od}[1]{\mathcal{O}(#1)} 
\nc{\kp}{\kappa}
\nc{\one}{\uppercase\expandafter{\romannumeral1}}
\nc{\two}{\uppercase\expandafter{\romannumeral2}}
\nc{\three}{\uppercase\expandafter{\romannumeral3}}
\def\({\left(}
\def\){\right)}
\def\[{\left[}
\def\]{\right]}
\def\e{\begin{equation}}
\def\q{\end{equation}}
\def\m{\begin{eqnarray}}
\def\n{\end{eqnarray}}
\nc{\Eq}[1]{Eq.~\eqref{#1}}     
\nc{\Fig}[1]{Fig.~\ref{#1}}     
\nc{\Table}[1]{Table~\ref{#1}}  
\nc{\Sec}[1]{Sec.~\ref{#1}}     
\nc{\Msun}{M_\odot}             
\nc{\fpbh}{f_{\mathrm{pbh}}}    
\nc{\fpbhn}{f_{\mathrm{pbh0}}}    
\nc{\mR}{\mathcal{R}} 
\nc{\seq}{\sigma_{\mathrm{eq}}}
\nc{\ogw}{\Omega_{\mathrm{GW}}}
\nc{\gpcyr}{\mathrm{Gpc}^{-3}\,\mathrm{yr}^{-1}}
\nc{\lvc}{LIGO/Virgo} 
\nc{\SNR}{\mathrm{SNR}} 
\nc{\mmin}{{m_{\mathrm{min}}}}
\nc{\mmax}{{m_{\mathrm{max}}}}
\nc{\Mmin}{{M_{\mathrm{min}}}}
\nc{\fmin}{{f_{\mathrm{min}}}}
\nc{\VT}{\mathrm{VT}}
\nc{\rhoGW}{\rho_{\mathrm{GW}}}
\nc{\vth}{\vec{\theta}}
\nc{\vd}{\vec{d}}
\nc{\vla}{\vec{\lambda}}
\nc{\Nobs}{N_{\mathrm{obs}}}
\nc{\av}[1]{\langle #1 \rangle} 
\nc{\km}{\mathrm{km}}
\nc{\Mpc}{\mathrm{Mpc}}
\nc{\Tobs}{T_{\mathrm{obs}}}
\nc{\Ntemp}{N_{\mathrm{temp}}}
\nc{\addref}{[\textcolor{red}{add ref}] } 
\nc{\eg}{\textit{e.g.~}}
\nc{\app}{\approx}
\nc{\hf}{\frac{1}{2}}
\nc{\discuss}{\textcolor{red}{Add discussion here!}}
\nc{\red}[1]{\textcolor{red}{#1}}
\nc{\mH}{\mathcal{H}}
\nc{\cs}{c_s^2}
\nc{\Sij}[1]{S_{ij}^{(#1)}}
\nc{\vi}[1]{v_i^{(#1)}}
\nc{\no}{\nonumber}
\def\<{\left\langle}
\def\>{\right\rangle}

\def\half{{1\over 2}}
\nc{\bk}{\bm{k}}
\nc{\bq}{\bm{q}}
\nc{\bp}{\bm{p}}
\nc{\bl}{\bm{l}}
\nc{\bx}{\bm{x}}
\nc{\be}{\mathbf{e}}
\nc{\mS}{\mathcal{S}}
\nc{\te}{\tilde{\eta}}
\nc{\tp}{\tilde{p}}
\nc{\tk}{\tilde{k}}
\nc{\tx}{\tilde{x}}
\nc{\tF}{\tilde{F}}
\nc{\tA}{\tilde{A}}
\nc{\mkpq}{|\bk-\bp-\bq|}
\nc{\mpq}{|\bp-\bq|}
\nc{\mkp}{|\bk-\bp|}
\nc{\mSi}[1]{\mS^{(#1)}({\bk, \eta})}
\nc{\vk}{\vec{k}}
\nc{\kstar}{k_*}
\nc{\xstar}{x_*}
\nc{\mpbh}{m_{\rm{pbh}}}
\nc{\Ci}{\mathrm{Ci}}
\nc{\Si}{\mathrm{Si}}
\nc{\fnl}{f_\mathrm{NL}}
\nc{\gnl}{g_\mathrm{NL}}
\nc{\Fnl}{F_\mathrm{NL}}
\nc{\Gnl}{G_\mathrm{NL}}
\nc{\togw}{\tilde{\Omega}}

\renewcommand{\vec}[1]{\boldsymbol{#1}} 

\begin{document}
	
\title{Gravitational waves induced by the local-type non-Gaussian curvature perturbations}
	
\author{Chen Yuan}
\email{yuanchen@itp.ac.cn}
\affiliation{CAS Key Laboratory of Theoretical Physics, 
Institute of Theoretical Physics, Chinese Academy of Sciences,
Beijing 100190, China}
\affiliation{School of Physical Sciences, 
University of Chinese Academy of Sciences, 
No. 19A Yuquan Road, Beijing 100049, China}
	
\author{Qing-Guo Huang}
\email{Corresponding author: huangqg@itp.ac.cn}
\affiliation{CAS Key Laboratory of Theoretical Physics, 
Institute of Theoretical Physics, Chinese Academy of Sciences,
Beijing 100190, China}
\affiliation{School of Physical Sciences, 
University of Chinese Academy of Sciences, 
No. 19A Yuquan Road, Beijing 100049, China}
\affiliation{School of Fundamental Physics and Mathematical Sciences
Hangzhou Institute for Advanced Study, UCAS, Hangzhou 310024, China}
\affiliation{Center for Gravitation and Cosmology, 
College of Physical Science and Technology, 
Yangzhou University, Yangzhou 225009, China}
\affiliation{Synergetic Innovation Center for Quantum Effects and Applications, Hunan Normal University, Changsha 410081, China}
	
\date{\today}

\begin{abstract}
 
The current observational constraints still leave a substantial mass window $\sim [10^{-16},10^{-14}] \cup [10^{-13},10^{-12}] M_\odot$ for primordial black holes (PBHs) representing all of dark matter (DM) in our Universe. The gravitational waves (GWs) induced by the curvature perturbations are inevitably generated during the formation of these PBHs, and fall in the frequency band of LISA. Such scalar induced gravitational waves (SIGWs) are supposed to be definitely detected by LISA even when the second-order local-type non-Gaussianity characterized by the parameter $F_{\rm NL}$ is taken into account. In this letter, we give a comprehensive analysis of the 
GWs induced by the local-type non-Gaussian curvature perturbations up to the third-order denoted by the non-linear parameter $G_{\rm NL}$, and we find that a log-dependent slope of SIGWs in the infrared region is generically predicted and the amplitude of SIGWs can be further suppressed by several orders of magnitude. As a result, the null-detection of SIGWs by LISA cannot rule out the possibility of PBHs making up all of DM.


\end{abstract}
	
	
\maketitle
	
{\it Introduction. } The nature of dark matter (DM) remains one of the most important problems in fundamental physics. The detection of gravitational waves (GWs) generated by the coalescence of two $\sim 30\Msun$ black holes \cite{Abbott:2016blz} has not only marked the beginning of GW astronomy but also renewed the postulation that primordial black holes (PBHs) may consist of all or some of DM in our Universe. 
PBHs are supposed to form from the collapse of over-dense regions in the radiation dominated (RD) era. An over-dense region is generated when very large curvature perturbation (for example, $\zeta>\zeta_c\simeq1$ \cite{Musco:2008hv,Musco:2004ak,Musco:2012au,Harada:2013epa}) re-enters the horizon, and all the matter inside the Hubble volume would collapse to form a PBH. The mass of a PBH is related to the comoving wavelength of the perturbation mode, $k_*$, by \cite{Hawking:1971ei,Carr:1974nx}, 
\e\label{mpbh}
{m_{\mathrm{pbh}}^*}\approx2.3\times10^{18}\Msun\left(\frac{3.91}{g_*^{\mathrm{form}}}\right)^{1 / 6}\left(\frac{H_{0}}{f_*}\right)^{2},
\q 
where $g_*^{\mathrm{form}}$ is the degrees of freedom at formation, $H_0$ is the Hubble constant and $k_*=2\pi f_*$. 
In the frequency band from $0.1$mHz to $100$mHz where LISA has a usable sensitivity, the corresponding PBH mass range are roughly from $8\times10^{-16}\Msun$ to $8\times10^{-10}\Msun$.
PBHs lighter than $\sim10^{-18}\Msun$ or heavier than $\sim10^3\Msun$ have been drastically constrained through Hawking radiation \cite{Carr:2009jm} and the accretion of primordial gas onto PBHs on CMB scales \cite{Chen:2016pud,Ali-Haimoud:2016mbv,Blum:2016cjs,Horowitz:2016lib} respectively. Although various investigations have put constraints on the fraction of PBHs in DM in the mass range $[10^{-12},10^{3}]\Msun$ to no more than a few parts in hundred \cite{Bird:2016dcv,Garcia-Bellido:2017fdg,Sasaki:2018dmp,Barack:2018yly,Chen:2018czv,Chen:2018rzo,Chen:2019irf,Barnacka:2012bm,Graham:2015apa,Niikura:2017zjd,Griest:2013esa,%
	Tisserand:2006zx,Brandt:2016aco,Gaggero:2016dpq,Niikura:2019kqi,%
	Wang:2016ana,Abbott:2018oah,Magee:2018opb,Chen:2019irf,Montero-Camacho:2019jte,Laha:2019ssq,Chen:2019xse},
the mass window $[10^{-16},10^{-14}] \cup [10^{-13},10^{-12}] M_\odot$ still opens for the possibility that all of DM are in the form of PBHs \cite{Defillon:2014wla,Katz:2018zrn}. See a recent review in \cite{Carr:2020gox}.

It is known that the scalar perturbations would source the gravitational waves at second-order \cite{tomita1967non,Matarrese:1992rp,Matarrese:1993zf,Matarrese:1997ay,Noh:2004bc,Carbone:2004iv,Nakamura:2004rm}, namely the scalar induced gravitational waves (SIGWs). SIGW inevitably generated during the formation of PBHs provides a new powerful tool to search for PBHs (in particular, those in the mass window $[10^{-16},10^{-14}] \cup [10^{-13},10^{-12}] M_\odot$ for constituting all of DM)  \cite{Saito:2008jc,Yuan:2019udt}, and the log-dependent slope of SIGWs in the infrared region near the peak is supposed to be a distinguishing feature from other astrophysical sources \cite{Yuan:2019wwo}. 
See some recent relevant works about SIGWs in \cite{Ananda:2006af,Baumann:2007zm,Assadullahi:2009jc,Bugaev:2009zh,Saito:2009jt,Bugaev:2010bb,Nakama:2016enz,Garcia-Bellido:2017aan,Espinosa:2018eve,Kohri:2018awv,Bartolo:2018evs,Bartolo:2018rku,Inomata:2018epa,Clesse:2018ogk,Cai:2019amo,Inomata:2019zqy,Inomata:2019ivs,Inomata:2019yww,Domenech:2019quo,Cai:2019elf,Cai:2019cdl,Tomikawa:2019tvi,DeLuca:2019ufz,Fu:2019vqc,Kawasaki:2019hvt,Lin:2020goi,Ota:2020vfn,Inomata:2020lmk,Giovannini:2020qta,Lu:2020diy,Domenech:2020kqm}.

Usually, the probability density function (PDF) of curvature perturbation is assumed to be Gaussian. However, PBHs are significantly generated by the enhanced scalar curvature perturbation over that measured by the cosmic microwave background on the very large scales, and the non-negligible non-Gaussianity is also expected to be generated. Since the formation of PBHs takes place at the tail of the PDF, the non-Gaussianity would drastically alter the PBH abundance \cite{Byrnes:2012yx}, and can suppress the SIGWs by several orders of magnitude \cite{Nakama:2016gzw}. Recently, the GWs induced by scalar curvature perturbation with a second-order (\eg $\Fnl$-order) local-type non-Gaussianity were studied in \cite{Cai:2018dig,Unal:2018yaa}. However, $\Fnl$ is just the coefficient of the even-order curvature perturbation, and it is necessary to extend to the odd-order curvature perturbation (\eg $\Gnl$-order) for achieving a general analysis of GWs induced by the local-type non-Gaussian scalar curvature perturbations. 

In this letter, we provide a comprehensive analysis of GWs induced by the local-type non-Gaussian curvature perturbations up to the third-order (\eg $\Gnl$-order). We find a universal $k^3\ln^2 (4k_*^2/3k^2)$ behavior of non-Gaussian contribution to SIGWs in the infrared region and the SIGWs can be undetectable by LISA once the $\Gnl$-order non-Gaussianity is taken into account. 
The goal of this work is to explore whether the null detection of GWs induced by non-Gaussian perturbations can rule out PBH DM or not. As a benchmark example, we consider a local-type model and treat the non-Gaussian parameters as free parameters.

{\it SIGWs induced by local-type non-Gaussian scalar curvature perturbations. } Let's start with the perturbed metric in Newton gauge 
\e
ds^2 = a^2\left\{-(1+2\phi)d\eta^2+[(1-2\phi)\delta_{ij}+\half h_{ij}]dx^i dx^j\right\},
\q
where $\phi$ is the linear scalar perturbation and $h_{ij}$ is the second-order tensor perturbation. The equation of motion for $h_{ij}$ is obtained by the second-order Einstein equation 
\e\label{eqh}
h_{i j}^{\prime \prime}+2 \mathcal{H} h_{i j}^{\prime}-\nabla^{2} h_{i j}=-4 \mathcal{T}_{i j}^{\ell m} S_{\ell m},
\q
where $\mH\equiv a'/a$ is the conformal Hubble parameter and the prime denotes derivative with respect to the conformal time $\eta$. The transverse and traceless projection operator in Fourier space is defined by 
\e
\mathcal{T}_{i j}^{\ell m}(\vec{k})\equiv 
e_{ij}^{(+)}(\vec{k})e^{(+)lm}(\vec{k})
+
e_{ij}^{(\times)}(\vec{k})e^{(\times)lm}(\vec{k}), \nonumber
\q
where the polarization tensors are $(e_i e_j - \bar{e}_i \bar{e}_j)/\sqrt{2}$ and $(e_i \bar{e}_j + \bar{e}_i e_j)/\sqrt{2}$ 
for $+$ and $\times$ modes respectively. From now on, we choose $e=(1,0,0)$ and $\bar{e}=(0,1,0)$.
During the RD era, the source term reads 
\m
S_{ij}\!\!=\!3\phi\p_i\p_j\phi\!-\!{1\over\mH}(\p_i\phi'\p_j\phi+\p_i\phi\p_j\phi')\!-\!{1\over\mH^2}\p_i\phi'\p_j\phi'\!.\quad
\n
The observational quantity, $\ogw(k,\eta)$, is defined as the energy density of GWs per logarithm wavelength normalized by the critical energy, $\rho_{c}(\eta)$, namely
\e\label{ogwd}
\ogw(k,\eta)\equiv{1 \over \rho_c(\eta)}{\mathrm{d}\rho_{\mathrm{GW}}(k,\eta) \over \mathrm{d} \ln k}
={k^3\over48\pi^2}\(k\over\mH\)^2\overline{\<|h_{\vec{k}}(\eta)|^2\>},
\q
where $\mathrm{d}\rho_{\mathrm{GW}}$ is the energy density of GWs in the frequency range $(k, k+\mathrm{d}k)$.
The GWs will be induced by curvature perturbations throughout the RD era, and the density parameter at the matter-radiation equality is $\ogw(k)=\ogw(k,\eta\rightarrow\infty)$.
After solving Eq.~(\ref{eqh}) by Green's function in Fourier space, $\ogw(k)$ can be expressed by, \cite{Kohri:2018awv}, 
\e\label{ogw}
\ogw(k)
=\int_0^\infty\mathrm{d}u\int_{|1-u|}^{1+u}\mathrm{d}v~P_\zeta(uk)P_\zeta(vk)
I(u,v),
\q
where the dimensionless curvature power spectrum $P_\zeta(k)$ is defined as
\m
\<\zeta(\vec{k})\zeta(\vec{k'})\>
&&\equiv {2\pi^2\over k^3}P_\zeta(k)\delta(\vec{k}+\vec{k'}).
\n
$\ogw(k)$ in Eq.~(\ref{ogw}) is valid during the RD era and the nowadays density parameter of SIGWs is given by $\Omega_{\mathrm{GW},0}(k) = \Omega_{r}\times\ogw(k)$ where $\Omega_{r}$ is the density parameter for radiation.
Here we use the the comoving curvature perturbation $\zeta$ which is related to $\phi$ by $\phi=(2/3)\zeta$, and $I(u,v)$ takes the form, \cite{Kohri:2018awv}, 
\m
I(u,v)&&={1\over12}\({4v^2-(1+v^2-u^2)^2 \over 4uv}\)^2\(3(u^2+v^2-3)\over 4u^3v^3\)^2\no\\
&&\times
\Bigg(
\Big(-4uv+(u^2+v^2-3)\ln\Big|{3-(u+v)^2\over 3-(u-v)^2}\Big|\Big)^2\no\\
&&+\pi^2\(u^2+v^2-3\)^2\Theta(u+v-\sqrt{3})
\Bigg).
\n

For the local-type non-Gaussian curvature perturbation, in real space, the curvature curvature perturbation $\zeta$ can be expanded in terms of the Gaussian part, $\zeta_g$, as follows 
\e
\zeta(\zeta_g) = \zeta_g +\Fnl\(\zeta_g^2-\<\zeta_g^2\>\)+\Gnl \zeta_g^3,
\q
where $\<\zeta_g^2\>=\int P_g(k)\mathrm{d}\ln k$ is the variance of the Gaussian part. Here, the non-Gaussian parameters $\Fnl$ and $\Gnl$ are related to $\fnl$ and $\gnl$ in literature by $\Fnl=3/5\fnl$ and $\Gnl=9/25\gnl$ respectively. For convenience, we define $P_g(k)\equiv A_g \tilde{P}_g(k)$ with $A_g$ standing for the dimensionless amplitude of the Gaussian part and $\tilde{P}_g(k)$ being normalized by $\int \tilde{P}_g(k)\mathrm{d}\ln k =1$.
Therefore, including the corrections from the non-Gaussian parts, the effective curvature power spectrum reads 
\m\label{ph}
{P}_{\zeta}(k)=\alpha_1 {\tilde P}_1(k)+\alpha_2 {\tilde P}_2(k)+\alpha_3 {\tilde P}_3(k),
\n
where $\alpha_1\equiv A_g\(1+3\Gnl A_g\)^2$, $\alpha_2\equiv \Fnl^2A_g^2$, $\alpha_3\equiv 3\Gnl^2A_g^3$, ${\tilde P}_1\equiv {\tilde P}_g$ and the non-Gaussian power spectra $\tilde{P}_2(k)$ and $\tilde{P}_3(k)$ read 
\m\label{p2}
{\tilde P}_2(k)&&\equiv \int \mathrm{d}^3 p{k^3 \tilde{P}_g(p)\tilde{P}_g(|\vec{k}-\vec{p}|)\over 2\pi p^3|\vec{k}-\vec{p}|^3}\no\\
&&=\int_{0}^{+\infty}\!\!\!\mathrm{d}v \!\int_{|1-v|}^{1+v}\mathrm{d}u~\frac{\tilde{P}_g(uk)\tilde{P}_g(vk)}{u^2 v^2},\\\label{p3}
{\tilde P}_3(k)&&\equiv \int \mathrm{d}^3 p\mathrm{d}^3 q{k^3\tilde{P}_g(p)\tilde{P}_g(q)\tilde{P}_g(|\vec{k}-\vec{p}-\vec{q}|)\over 8 \pi^2 p^3 q^3 |\vec{k}-\vec{p}-\vec{q}|^3}\no\\
&&=\int_{0}^{+\infty}\!\!\!\mathrm{d}v \!\int_{|1-v|}^{1+v}\mathrm{d}u~\frac{\tilde{P}_g(uk)\tilde{P}_2(vk)}{2u^2 v^2}. 
\n
Here we change the variables to $|\vec{k}-\vec{p}|/k=u$ and $p/k=v$. And then the density parameter $\ogw(k)$ can be written by 
\m\label{ogw1}
\ogw (k)=\sum_{i,j=1}^3 \alpha_i \alpha_j {\tilde \Omega}_{ij}(k), 
\n
where 
\m\label{ogwij}
{\tilde \Omega}_{ij}(k)=\int_0^\infty
\mathrm{d}u
\int_{|1-u|}^{1+u}\!\!\mathrm{d}v
{\tilde P}_i(uk) {\tilde P}_j(vk) I(u,v). 
\n
By our definition, ${\tilde \Omega}_{ij}={\tilde \Omega}_{ji}$.
Roughly speaking, the amplitudes of shape functions ${\tilde \Omega}_{ij}$ are ${\cal O}(1)$ because ${\tilde P}_g$ is normalized to be unity, and thus the amplitude of $\ogw(k)$ can be estimated by the dominated coefficient among $\alpha_i\alpha_j$. 

\begin{figure}[htbp!]
	\centering
	\includegraphics[width = 0.45\textwidth]{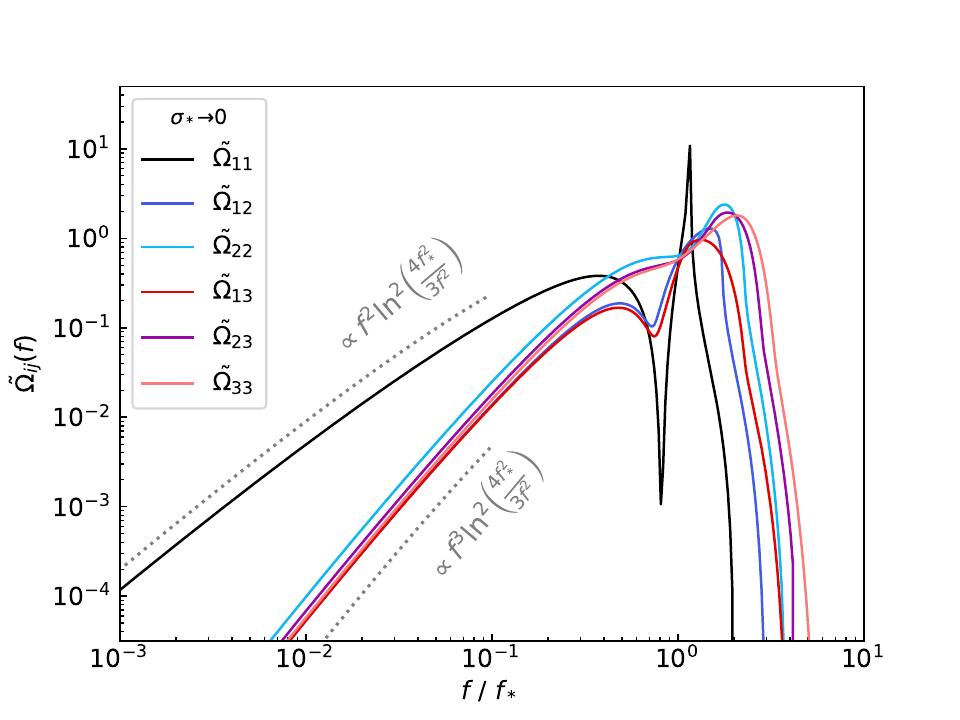}
	\includegraphics[width = 0.45\textwidth]{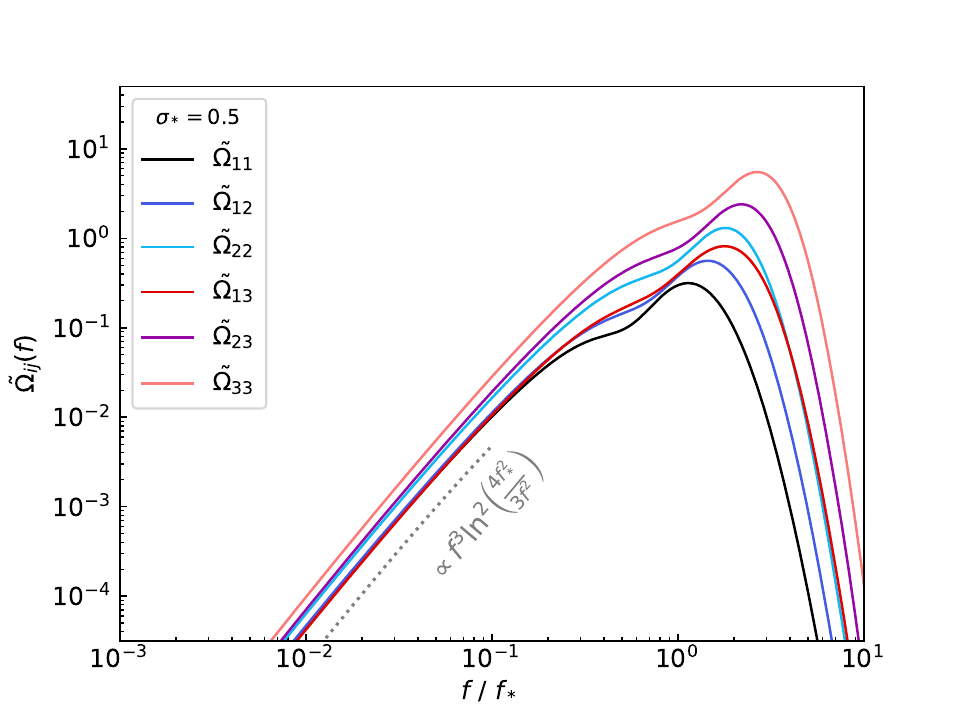}
	\caption{\label{togw} 
		The shapes of $\togw_{ij}$ for the log-normal power spectrum. The upper panel corresponds to $\sigma_*\rightarrow 0$ and the lower panel correspond to $\sigma_*=0.5$. Two reference lines of $f^2\ln^2 (4f_*^2/3f^2)$ and $f^3\ln^2 (4f_*^2/3f^2)$ are also illustrated, corresponding to slopes of $2-2/\ln(2f_*/\sqrt{3}f)$ and $3-2/\ln(2f_*/\sqrt{3}f)$ respectively.
	}
\end{figure}

It is difficult to obtain an analytic result of ${\tilde \Omega}_{ij}$ for a general curvature power spectrum. Here we consider a log-normal power spectrum for the Gaussian part, namely 
\e\label{ln}
{\tilde P}_1(k)={A_g\over \sqrt{2\pi\sigma_*^2}}\exp\(-\frac{\ln(k/k_*)^2}{2\sigma_*^2}\),
\q
to illustrate the main features of SIGWs. In the limit of $\sigma_*\rightarrow 0$, $\tilde{P}_1(k)$ approaches to a $\delta$ power spectrum, namely $k_*\delta (k-k_*)$. In such limit, the higher order power spectra become 
\m
\tilde{P}_2(k) && = \tk^2\Theta(2-\tk)\\
\tilde{P}_3(k) && = M(\tk)\Theta(3-\tk), 
\n
where $\tk\equiv k/k_*$, $M(\tk)\equiv \min(2\tk^3,(3-\tk)\tk^2)$ and $\Theta$ is the Heaviside theta function. Then ${\tilde \Omega}_{ij}$ becomes 
\m\label{ngeq}
\tilde{\Omega}_{11}
&&={\tk^{-2}}I\({1\over \tk},{1\over \tk}\)\Theta(2-\tk), \no\\
\tilde{\Omega}_{12}
&&=\tk\int_{\big|1-{1\over \tk}\big|}^{\min({2\over \tk},1+{1\over \tk})}\mathrm{d}v~v^2I\({1\over \tk},v\)\Theta(3-\tk), \no\\
\tilde{\Omega}_{22}
&&=\tk^4\int_{0}^{{2\over \tk}}\mathrm{d}v\int_{|1-v|}^{1+v}\mathrm{d}u~u^2v^2I(u,v)\Theta(2-u\tk), \no\\
\tilde{\Omega}_{13}
&&=\tk^{-1}\int_{\big|1-{1\over \tk}\big|}^{\min({3\over \tk},1+{1\over \tk})}\mathrm{d}v~M(v\tk)I\({1\over \tk},v\)\Theta(4-\tk), \no\\
\tilde{\Omega}_{23}
&&=\tk^2\int_{0}^{3\over\tk}\mathrm{d}v\int_{|1-v|}^{1+v}\mathrm{d}u~u^2I(u,v) M(v\tk)\Theta(2-u\tk), \no\\
\tilde{\Omega}_{33}
&&=\int_{0}^{{3\over \tk}}\mathrm{d}v\int_{|1-v|}^{1+v}\mathrm{d}u~M(u\tk)M(v\tk)I(u,v)\Theta(3-u\tk). \no
\n 
For a more general investigation, we plot the shape functions ${\tilde \Omega}_{ij}(k)$ in Fig.~\ref{togw} for the log-normal power spectra with $\sigma_*\rightarrow 0$ (upper panel) and $\sigma_*=0.5$ (lower panel), respectively. Here ${\tilde \Omega}_{11}(k)$ is contributed by the Gaussian part, $\tilde{P}_1(k)$, and the others ${\tilde \Omega}_{ij}(k)$ are related to the non-Gaussian parts. 
From Fig.~\ref{togw}, the amplitudes of ${\tilde \Omega}_{ij}(k)$ is ${\cal O}(1)$ which is consistent with our previous estimation. In the infrared region $(k< k_*)$, the Gaussian contribution goes like $\togw_{11}(f)\propto f^2\ln^2(4f_*^2/3f^2)$ for a infinite narrow ($\sigma_*\to0$) power spectrum and $\togw_{11} (f)\propto f^3\ln^2(4f_*^2/3f^2)$ for $\sigma_*=0.5$, and the others ${\tilde \Omega}_{ij}(k)$ contributed by non-Gaussian parts goes like $\togw_{ij} (f)\propto f^3\ln^2(4f_*^2/3f^2)$ which are independent of $\sigma_*$. It indicates that the non-Gaussian contributions to the shape functions have a universal behavior with slope $n_{\rm GW}\equiv \frac{{\rm d}\ln\tilde{\Omega}_{ij}}{{\rm d}\ln f }=3-2/\ln(2f_*/\sqrt{3}f)$, 
and such a difference can be probably distinguished by LISA at high confidence level \cite{Yuan:2019wwo}. Even though $n_{\rm GW}\rightarrow 3$ in the far infrared limit ($k\ll k_*$) except ${\tilde \Omega}_{11}$ for a very narrow power spectrum, the SIGWs in the far infrared limit should be extremely suppressed and have much less opportunities to be detected. In this sense, such a log-dependent slope given in \cite{Yuan:2019wwo} is also reliable even for the non-Gaussian case, and can be taken as a generic feature for SIGWs accompanying the formation of PBHs. 



\begin{figure}[htbp!]
	\centering
	\includegraphics[width = 0.41\textwidth]{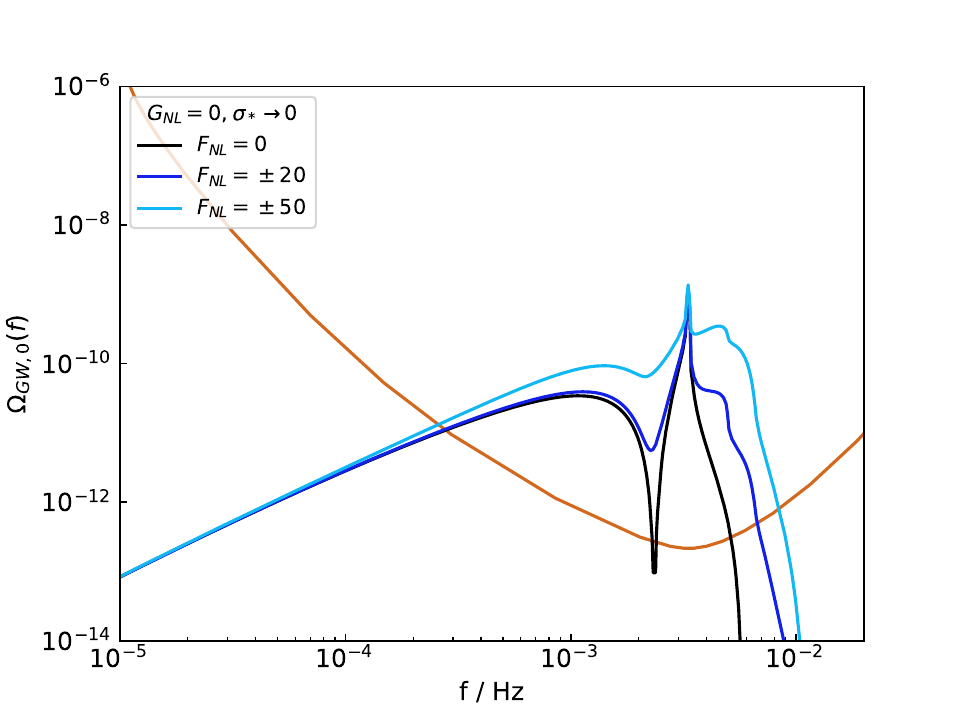}
	\includegraphics[width = 0.41\textwidth]{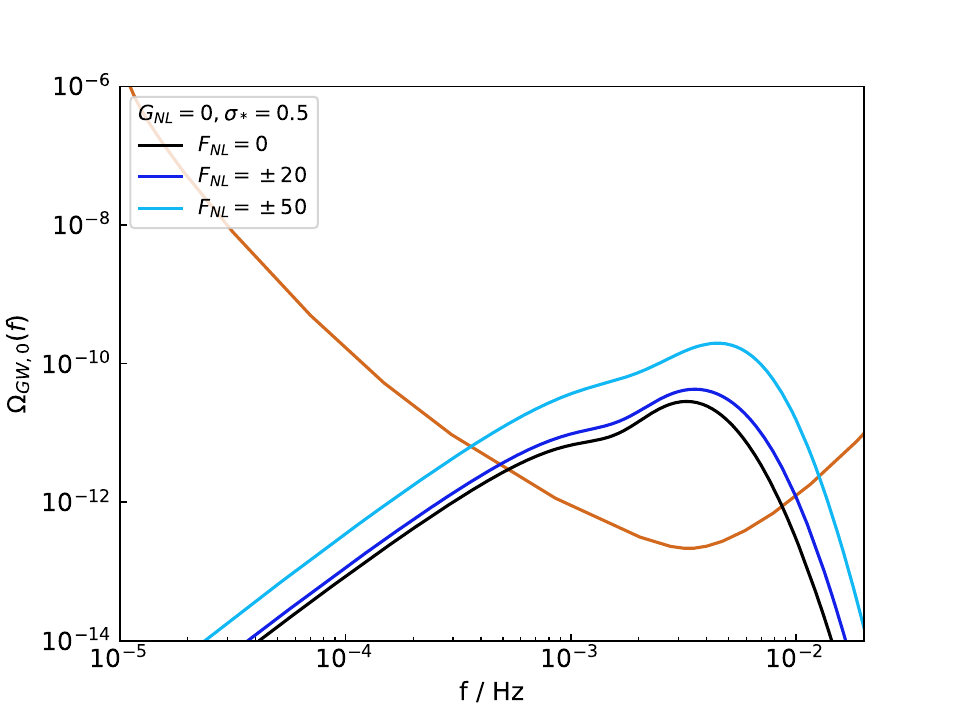}
	\includegraphics[width = 0.41\textwidth]{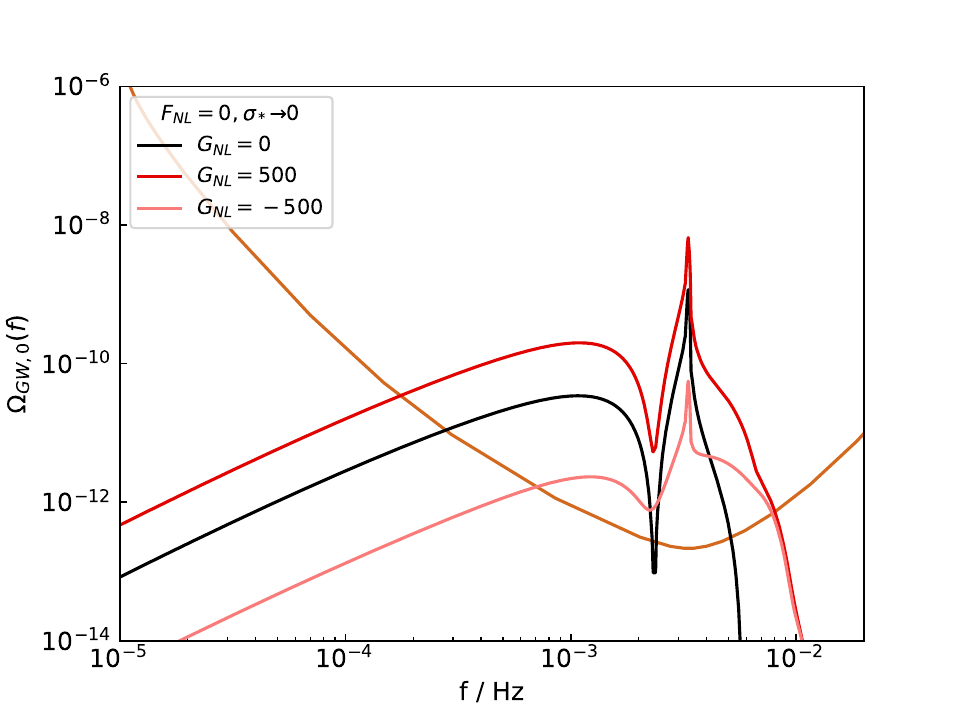}
	\includegraphics[width = 0.41\textwidth]{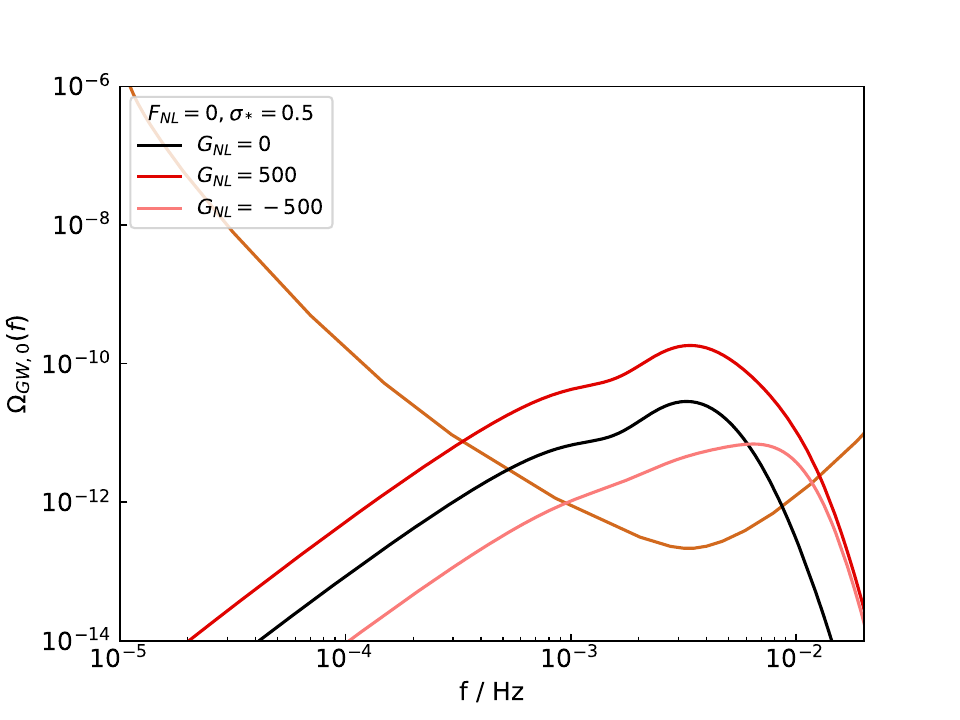}
	\caption{\label{ogwfnl}\label{ogwgnl} 
		The density parameter $\Omega_{\mathrm{GW},0}(f)$ induced by non-Gaussian curvature perturbations. Here we fix the amplitude of Gaussian spectrum to be $A_g=10^{-3}$ for $\mpbh^*=10^{-12}\Msun$. The brown line is the power-law integrated sensitivity curves \cite{Thrane:2013oya} for LISA, assuming a four year observation.
	}
\end{figure}

Keeping $A_g=10^{-3}$ fixed, for $m_{\rm pbh}^*=10^{-12}M_\odot$, the density parameter $\Omega_{\mathrm{GW},0}(f)$ with non-Gaussianity is illustrated in Fig.~\ref{ogwfnl} for some typical values of $\Fnl$ and $\Gnl$. Since $\Fnl$ contributes to $\Omega_{\mathrm{GW},0}(f)$ in the form of $\alpha_2=\Fnl^2 A_g^2$, $\Omega_{\mathrm{GW},0}(f)$ is only related to the absolute value of $\Fnl$. In contrast, the contribution to $\Omega_{\mathrm{GW},0}(f)$ from $\Gnl$ encodes in terms of $\alpha_1=A_g\(1+3\Gnl A_g\)^2$ and $\alpha_3=3\Gnl^2A_g^3$, and then a negative value of $\Gnl$ can suppress $\Omega_{\mathrm{GW},0}(f)$ compared to a positive $\Gnl$. 

{\it Testing the postulation of PBH DM by LISA. } Similar to \cite{Cai:2018dig}, we also consider a narrow curvature power spectrum, and then the PBHs are roughly monochromatic. According to \cite{Byrnes:2012yx}, the formation probability $\beta$ of PBHs can be evaluated by 
\m\label{beta}
\beta= \int_{\zeta(\zeta_g)>\zeta_c}  {1\over \sqrt{2\pi A_g}}\exp\(-{\zeta_g^2\over 2A_g}\)\mathrm{d}\zeta_g.
\n
Here, the integrating regions include all of the regions satisfying $\zeta(\zeta_g)>\zeta_c$.
And the formation probability $\beta$ is related to the abundance of PBHs in DM by, \cite{Nakama:2016gzw}, 
\m\label{fpbh}
f_{\mathrm{pbh}}\simeq 2.5 \times 10^{8}\beta\left(\frac{g_*^{\mathrm{form}}}{10.75}\right)^{-\frac{1}{4}}\left(\frac{m_{\mathrm{pbh}}}{M_{\odot}}\right)^{-\frac{1}{2}}.
\n

\begin{figure}[htbp!]
	\centering
	\includegraphics[width = 0.5\textwidth]{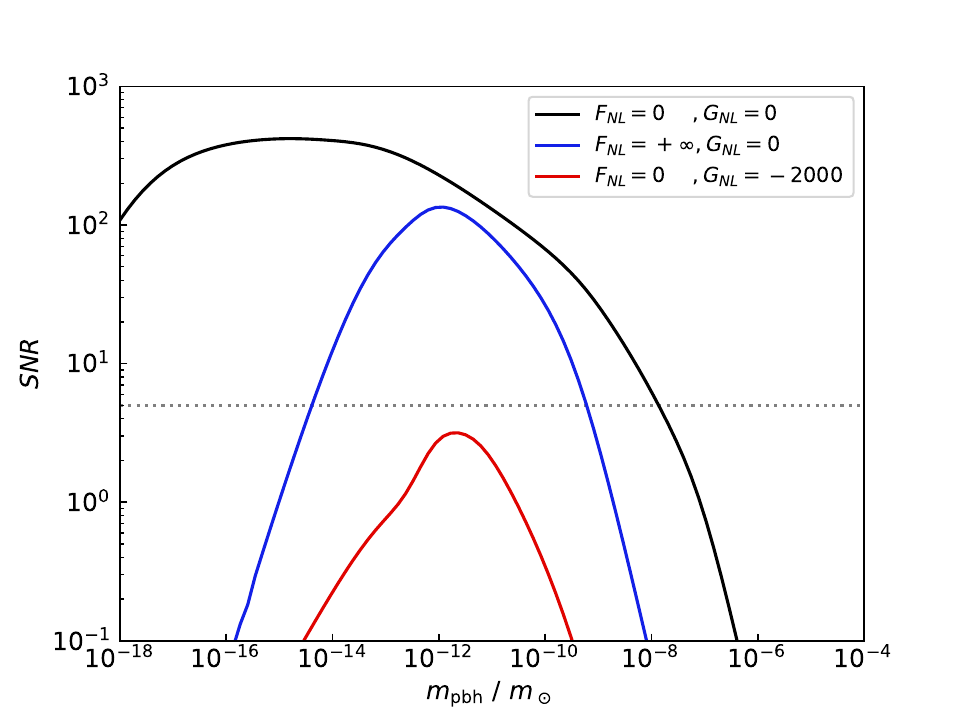}
	\caption{\label{SNR} 
		The expected SNR of SIGWs generated by monochromatic PBHs as a function of $\mpbh$. We keep $\fpbh=1$ fixed and assume a four year observation time for LISA. A reference line, $\rho=5$, is drawn for comparison.
	}
\end{figure}

\begin{figure}[htbp!]
	\centering
	\includegraphics[width = 0.5\textwidth]{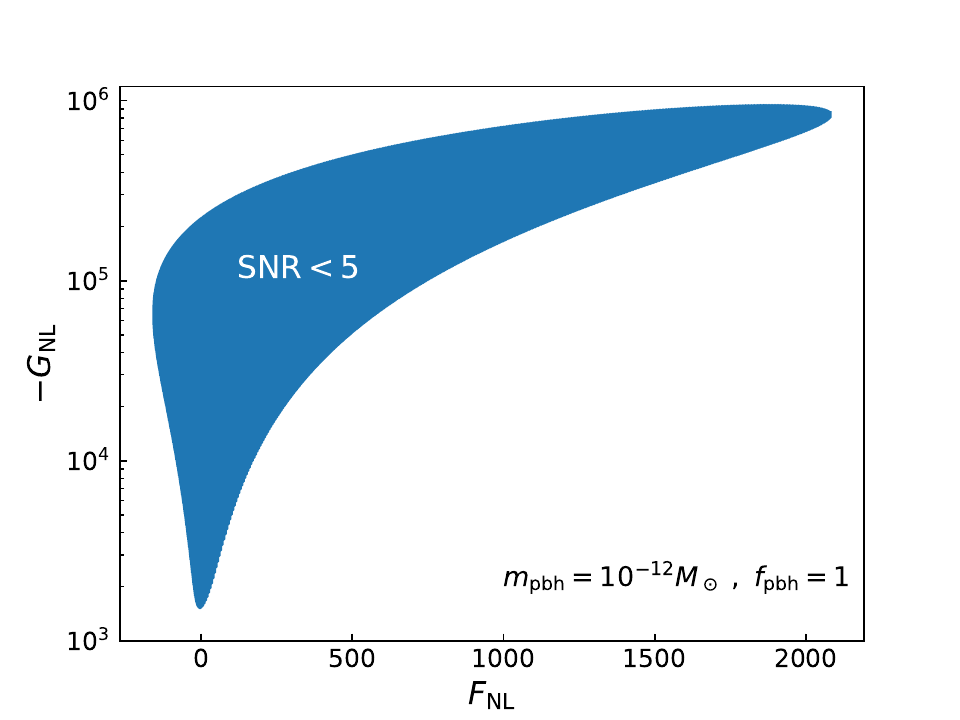}
	\caption{\label{fgsnr} 
		The allowed parameter space for $\Fnl$ and $\Gnl$ with SNR$<5$, for example, for $\mpbh=10^{-12} M_\odot$ and $\fpbh=1$.  
	}
\end{figure}

To quantitatively estimate the detectability of the SIGW signals by LISA, we evaluate the expected signal-to-noise ratio (SNR) ($\rho$). The SNR of detecting a GW background is given by,  \cite{Thrane:2013oya,Allen:1997ad}, 
\m
\rho^2 \! = \! T \!\!  \int  \!\! \mathrm{d}f
\frac{R(f)^2S_h(f)^2}
{[1+
	R(f)^2] S_h(f)^2 + 
	P_n(f)^2+2S_h(f)P_n(f)},\no\\
\n
where $S_h(f) = 3H_0^2\Omega_{\mathrm{GW},0}/(2\pi^2 f^3)$. $R(f)$ and $P_n(f)$ are the signal transfer function and the total noise spectrum respectively \cite{Cornish:2018dyw}. In order to test the postulation of PBH DM, we assume that PBHs constitute all of DM in our Universe, \eg $f_{\rm pbh}=1$, and plot the expected SNR for a different mass of PBHs in Fig.~\ref{SNR} where a dotted line corresponds to $\rho=5$ beyond which the GW signals are supposed to be distinguished from the noise. Here we focus on PBHs with mass heavier than $10^{-18}M_\odot$, otherwise, they would have been evaporated completely by Hawking radiations. 
For the absence of non-Gaussianity, LISA can detect SIGWs accompanying the formation of PBHs in the mass range $\sim [10^{-18},1.4\times 10^{-8}]\Msun$. 
Taking into account the non-Gaussianity up to the $\Fnl$-order, the amplitude of $\Omega_{\mathrm{GW},0}$ will be mostly suppressed in the limit of $\Fnl=+\infty$. 
Although $\Omega_{\mathrm{GW},0}$ is suppressed by several orders of magnitude in such limit, it is still optimistic that LISA can detect the SIGWs for PBHs in the mass range $\sim[4.1\times 10^{-15}, 6.0\times 10^{-10}]\Msun$.
However, from Fig.~\ref{SNR}, we see that the SNR is always less than 5 for PBHs with arbitrary mass if, for example, $\Fnl=0$ and $\Gnl=-2000$. Moreover, one can check that both $\Fnl$ and $\Gnl$ can span several orders of magnitude for $f_{\rm pbh}=1$ and SNR$<$5. For example, the allowed parameter space for $\Fnl$ and $\Gnl$ with SNR$<$5 is given in Fig.~\ref{fgsnr} for $\mpbh=10^{-12} M_\odot$ and $\fpbh=1$. 
Therefore, a null detection of SIGWs can not rule out the possibility of PBH constituting all of DM in the remaining mass window $\sim [10^{-16},10^{-14}] \cup [10^{-13},10^{-12}] M_\odot$, but may leave a possibility for large primordial non-Gaussianity at the third order.

{\it Conclusion. } In this letter, we provide a comprehensive analysis of the GWs induced by the local-type non-Gaussian curvature perturbations up to the third-order (\eg the $\Gnl$ order). We find that in general the SIGWs can be written in the form of Eq.~(\ref{ogw1}) where the amplitude of $\Omega_{\mathrm{GW},0}(f)$ is mainly determined by the dominated coefficient among $\alpha_i \alpha_j$, and the shape functions from the non-Gaussian parts have a universal behavior with slope $n_{\rm GW}=3-2/\ln(2f_*/\sqrt{3}f)$ in the infrared region. Even though a log-dependent slope of SIGW is proposed for a Gaussian curvature perturbation in \cite{Yuan:2019wwo}, we find such a log-dependent slope is quite generic even for the non-Gaussian curvature perturbations and can be taken as a generic feature for the SIGWs accompanying formation of PBHs. 


Because the effects on the SIGWs from non-Gaussianity are completely encoded in the modification of scalar power spectrum (\eg Eq.~(\ref{ph})), one can not identify any features in SIGWs as smoking-guns for non-Gaussianity. However, even though different models with or without non-Gaussianity in the very early Universe can produce the same SIGWs if they generate the same effective scalar power spectrum, the PDFs of the curvature perturbations are different, and then lead to different abundances of PBHs due to the different integration region in Eq.~(\ref{beta}). In this sense, we can identify the non-Gaussianity by considering both the SIGWs and the abundance of PBHs, not SIGWs only. 

In addition, we also explore the detectability of the GWs induced by the local-type non-Gaussian curvature perturbation for PBHs serving as all of DM by LISA in detail. Since the third-order non-Gaussianity can further suppress the amplitude of $\Omega_{\mathrm{GW},0}(f)$, the SIGWs may not be detected by LISA even for PBHs in the mass range $\sim[10^{-16},10^{-14}] \cup [10^{-13},10^{-12}] M_\odot$ making up all of the DM in our Universe. In this sense, a more sensitive GW detector than LISA is needed for testing the postulation of PBH DM. 

Although higher-order non-Gaussianities could either enhance or suppress the SIGWs, the $g_{NL}$-order has already made the SIGWs low enough to escape the detection of LISA, and thus a null detection by LISA cannot rule out the PBH DM hypothesis. As a result, we don't consider the higher-order non-Gaussianities in this letter.

To include the entire non-Gaussian effects instead of taking a local-type model, one can calculate the full probability density function to evaluate the formation of PBHs and the SIGWs based on \cite{Celoria:2021vjw}. Moreover, it has been proposed recently that the quantum diffusion effects during the formation of PBHs are not negligible \cite{Pattison:2017mbe,Ezquiaga:2019ftu,Ando:2020fjm,Pattison:2021oen}, and its correction to the formation of PBHs and SIGWs are worth considering. In addition, the connected diagrams also make contributions to SIGWs \cite{Unal:2018yaa,Adshead:2021hnm}, and should be taken into account for a full analysis. We will leave these for future work.


{\it Acknowledgments.} 
We would like to thank Zu-Cheng Chen and Shi Pi for useful conversations. 
Cosmological perturbations are derived using the \texttt{xPand} \cite{Pitrou:2013hga} package. We also acknowledge the use of HPC Cluster of ITP-CAS. 
This work is supported by the National Key Research and Development Program of China Grant No.2020YFC2201502, grants from NSFC (grant No. 11975019, 11690021, 11991052, 12047503), the Strategic Priority Research Program of Chinese Academy of Sciences (Grant No. XDB23000000, XDA15020701), and Key Research Program of Frontier Sciences, CAS, Grant NO. ZDBS-LY-7009.     

\bibliography{./ref}


\end{document}